\DocumentMetadata{}
\documentclass[manuscript,screen,acmsmall]{acmart}
\AtBeginDocument{%
  }

\setcopyright{acmcopyright}
\copyrightyear{2025}
\acmYear{2025}
\acmDOI{XXXXXXX.XXXXXXX}

\acmVolume{XX}
\acmNumber{XX}
\acmArticle{XX}
\acmMonth{XX}
\acmConference[CSCW '25]{Conference on Computer Supported Cooperative Work}{2025}{TBD}
\acmISBN{978-1-4503-XXXX-X/18/06}

\begin{document}

\title{``We’re losing our neighborhoods.
We’re losing our community'': A comparative analysis of community discourse in online and offline public spheres}

\author{Casey Randazzo}
\affiliation{%
  \institution{Rutgers University}
  \city{New Brunswick,NJ}
  \country{US}}
\email{cer124@rutgers.edu}

\author{Minkyung Kim}
\affiliation{%
  \institution{University of Illinois, Urbana-Champaign}
  \city{Urbana, Illinois}
  \country{US}}
\email{mkk84@illinois.edu}

\author{Melanie Kwestel}
\affiliation{%
  \institution{Rutgers University}
  \city{New Brunswick,NJ}
  \country{US}}
\email{melanie.kwestel@rutgers.edu}

\author{Marya L Doerfel}
\affiliation{%
  \institution{Rutgers University}
  \city{New Brunswick,NJ}
  \country{US}}
\email{mdoerfel@rutgers.edu}

\author{Tawfiq Ammari}
\affiliation{%
  \institution{Rutgers University}
  \city{New Brunswick,NJ}
  \country{US}}
\email{tawfiq.ammari@rutgers.edu}



\begin{abstract}
Recovering from crises, such as hurricanes or wildfires, is a complex process that can take weeks, months, or even decades to overcome. Crises have both acute (immediate) and chronic (long-term) effects on communities. Crisis informatics research often focuses on the immediate response phase of disasters, thereby overlooking the long-term recovery phase, which is critical for understanding the information needs of users undergoing challenges like climate gentrification and housing inequity. We fill this gap by investigating community discourse over eight months following Hurricane Ida in an online neighborhood Facebook group and Town Hall Meetings of a borough in the New York Metropolitan region. Using a mixed methods approach, we examined the use of social media to manage long-term disaster recovery. The findings revealed a significant overlap in topics, underscoring the interconnected nature of online and offline community discourse, and illuminated themes related to the long-term consequences of disasters. We conclude with recommendations aimed at helping designers and government leaders enhance participation across community forums and support recovery in the aftermath of disasters.
\end{abstract}





\received{1 July 2024}
\maketitle

\section{INTRODUCTION}
Recovering from a crisis—whether from extreme flooding due to a hurricane or the extensive damage caused by a wildfire—can be a long and arduous process that spans months, years, or decades \cite{dailey2020social, tierney2020social}. Yet crisis informatics, the study of how technological tools can aid in recovery \cite{starbirdintelligence}, predominantly focuses on the emergency response phase of disaster which occurs during or immediately following a crisis \cite{dailey2020social, soden2018informating}.\footnote{Disasters are large-scale catastrophes that introduce crises—unpredictable events or situations that generate uncertainty and disrupt objectives due to the urgent needs they create \cite{seeger1998communication}.} 

Temporal models in the crisis literature organize phases around a disrupting event with a concern for protecting reputation and anticipated threats \cite{coombs2020conceptualizing}. Other phase models emphasize the role leaders play in protecting the integrity of operations \cite{williams2017organizational}. Such views isolate sequential tasks into categories such as (a) signal detection, (b) preparation/ prevention, (c) containment/damage control, (d) business recovery, and (e) learning \cite{williams2017organizational}. Planned structures (e.g., policy, formal roles) or improvisation may be taken up at various points before, during, or after disruptive events. However, these categories lack clarity about when and how levels may play more prominent or background roles. Examining resilience across levels and over time, we are thus guided by the following research questions: (1) Over time, how are individual, organizational, and inter-organizational levels enacted and interdependent before, during, and in the aftermath of a crisis?

Although critical, too much focus on the emergency response phase can overshadow the equally important information needs of users during the long-term recovery phase. Individuals' long-term needs often revolve around challenges such as inequitable access to safe and affordable housing \cite{keenan2018climate}. For example, in the aftermath of a crisis, property prices can fluctuate based on their capacity to accommodate rising sea levels. This volatility in the market can become a source of opportunity for developers to replace affordable housing with high-end rentals \cite{keenan2018climate}. Climate gentrification is just one example of the often missed issues that can emerge months after a crisis \cite{tierney2020social, keenan2018climate}. Not considering the longer-term recovery phase of a crisis leaves a significant gap in our understanding of how communities use social media to create, share, and utilize information during and after disasters \cite{dailey2020social, soden2018informating}.

To address this gap, this study follows Dailey’s \cite[P.21]{dailey2020social} recommendation to extend ``the scope of analysis beyond the emergency period.'' As the author explains, this approach allows researchers to conduct ``the more challenging and often essential work of linking an emergency to what…follows from it'' \cite[P.21]{dailey2020social}. Thus, this study utilizes data following eight months after a disaster, reflecting a range of discussions about the short-term (e.g., search and rescue, securing shelter) and long-term consequences (e.g., mental health risks, displacement of large groups) of environmental disasters \cite{tierney2020social}. Specifically, we focused on a borough in the New York Metropolitan Region that was devastated by Hurricane Ida, a Category 4 storm that caused catastrophic flooding and tornado destruction, making the hurricane one of the most costly in U.S. history \cite{woodida, mckinleyida}.

Prior work has also been limited in the consideration of \textit{locality}, which is the local context or environment in which the technology is being used \cite{palen2018social, palen2016crisis}. Palen and Hughes \cite[P.507]{palen2018social} refer to this as a ``flattening of communication medium'' which can lead scholars being overly concerned with social media platforms as opposed to other information sources in a local community \cite{dailey2020social}. Semaan et al. \cite{Semaan2015public} argue that sociotechnical theorists and researchers should not confine the \textit{public sphere}, defined as communicative realms in which people discuss issues \cite{habermas1989jurgen}, to a single space. Thus, to better consider the locality of discourse after crises, we need ``a detailed analysis of the intersection of, and continuity between, digital and physical spaces'' \cite[P.7]{willems2019politics}. Locality does not cease to exist because of technological advancements \cite{hepp2009transculturality}.\footnote{In this study, technological advancements refers to the expansion of the Internet and digital networks that facilitate the distribution, production, and consumption of information \cite{hepp2009transculturality}.} Online communities, though seemingly detached from physical locations, can be deeply rooted in and reflective of material identities, enabling the emergence of translocal communities \cite{hepp2009transculturality}. Translocal communities are defined as locally anchored spaces that simultaneously transcend communication boundaries \cite[P.2]{waldherr2023translocal}. Examples of translocal communities include \textit{Online Neighborhood Group s}, which are digital interest or topic communities tied to geographic regions \cite{kim1999news, la1998social, weatherford1982interpersonal}. Such networks, often hosted on social media platforms like Facebook or Nextdoor \cite{aubin2024not, choksi2024under}, contain informal discussions about community issues \cite{carpini2004public}. In this way, Online Neighborhood Group s can be considered publics \cite{habermas1989jurgen}.

To follow Semaan et al.'s \cite{Semaan2015public} call and not confine publics it a digital space, we incorporate data from two community settings: a community's Online Neighborhood Group , hosted on Facebook, and transcripts from their in-person Town Hall Meetings. We then applied topic modeling to analyze the discussions from both Town Hall Meetings and Online Neighborhood Group s. Findings revealed a wide range of topics in both settings, exemplifying the complexity of community issues across the eight months following a disaster. In the Online Neighborhood Group , users shared information and support. Meanwhile, Town Hall Meetings offered a structured setting for addressing more direct community needs and providing feedback to local authorities. Next, we measured the similarities between online and offline contexts using Jaccard coefficients. The analysis revealed a moderate to high degree of similarity across most topics, providing empirical evidence for prior theoretical research on the co-constitutive nature of online and offline settings \cite{waldherr2021spaces, laurell2017spatial, keinert2021relational}. We also suggest several design recommendations aimed at enhancing participation both in physical Town Hall Meetings and through Online Neighborhood Group s that can be utilized by both designers and government officials.

\section{PRIOR WORK}
In this section, we start by reviewing literature about discourse in communities, especially during crisis events. We then summarize literature on public sphere theory. Finally, we discuss materiality in different modalities of community discourse.

\subsection{Community Discourse During Crises} \label{related_work_sec:comm_discourse_crises}
This study focuses on \textit{community discourse} during a disaster. Language—whether in words, sentences, or groups of utterances—constructs social contexts \cite{schiffrin1987discourse, howarth2000discourse}. In other words, regardless of what exists cognitively \cite{potter1987discourse}, discourse reflects what is being emphasized in communication by its participants. Qualitatively analyzing discourse can highlight how societal narratives and structures evolve to establish norms and ideologies \cite{giddens1979central, giddens1985time}. When examining textual data, a discourse approach allows scholars to view communication as more than information exchange or diffusion \cite{matassi2023know}. Instead, a discourse lens centers the meaning and interpretation of texts and language, supporting the evaluation of social media as a community discourse platform and cross-cultural dialogue \cite{papacharissi2009virtual, berger2012makes}. Discourse can also act as a resource upon which an actor uses to structure knowledge and make sense of a situation \cite{hardy2005discourse}.

Community discourse is often examined in \textit{community computing} studies, defined as technologically-enabled interactions ``among and between residents, organizations, government and businesses in a geographically bounded setting for local purposes and activities'' \cite[P.1]{kavanaugh2005community}. The phenomenon of talking to one’s neighbors online is referred to by various names in the literature, including online neighborhood networks (ONNs; \cite{meulenaere2021neighborhood}), online neighborhood social networks (ONSNs; \cite{vogel2021designing, vogel2021older}), neighborhood storytelling networks (e.g., \cite{jung2013delineation}), neighborhood hotspots \cite{meulenaere2021neighborhood}, neighborhood-related social media \cite{gatti2022ubiquitous}, online neighborhood forums or platforms \cite{afzalan2017creating, li2014living}, digital neighborhoods or publics \cite{bingham2017imagined, Smite_2013, willis2017digital}, local community platforms \cite{holzer2023towards}, hyperlocal online forums \cite{lopezeneighbors2017}, and community informatics \cite{shin2012community}. For simplicity, we will refer to these forums as Online Neighborhood Group s throughout the paper.

During crisis events, Online Neighborhood Group s have been touted for helping digital volunteers organize relief efforts \cite{starbirdintelligence, qu2009online}. In disaster contexts, Online Neighborhood Group s have been found to enable residents to engage in dialogue about a crisis in real-time \cite{bortree2009dialogic, seltzer2007dialogic, taylor2005diffusion}. In these groups, individuals find and share information pertaining to a crisis \cite{lindsay2011social, kim2018social} and organize others to help with volunteer activities like cleanup efforts \cite{velev2018information}. Online Neighborhood Group s have been found to contribute to the overall resilience capacity of communities \cite{aydin2018framework} as they can help enable the maintenance and formation of social capital (e.g., resources embedded in networks of social relationships; \cite{bourdieu1986force, coleman1988social, lin1999social}). Crisis informatic studies often focus on how to monitor information in digital settings and share that information with government or nonprofit organizations \cite{cobb2014designing, rodriguez2018handbook}. 

During extreme weather events, local governments communicate emergency and disaster management messages through various channels, including Town Hall Meetings. These messages function to mitigate risks, prepare citizens, and support recovery efforts \cite{drabek1991emergency}. Government leaders representing local, state, or federal levels ``have the potential to either mitigate or exacerbate the impact disaster has on the citizens they represent, how they make sense of, interpret, and reframe disaster has serious implications for how victims experience it'' \cite[P.209]{dennis2006making}. Ineffective communication in the aftermath of disasters can have consequences as residents expect government representatives to protect their safety \cite{boin2003public}. Taken together, such studies highlight how community discourse can construct social contexts in the event of disaster. In the following section, we discuss \textit{public sphere theory} to consider how online and offline platforms can emerge as crucial spaces for driving collective action around the short-term \textit{and} long-term consequences of disasters.

\subsubsection{Community organizing}
\label{subsubsec:comm_org}
Community organizing refers to the collective action of local residents addressing issues that threaten their community, such as public safety, housing, or disaster recovery \cite{christens2015community}. It typically serves two core purposes: fostering collective recovery and advocating for policy changes on local issues \cite{speer2003intentional,christens2015community}. Typically, organizing efforts begin informally through residents, nonprofits, and local groups, evolving into structured, institutional-level collaborations spanning multiple levels \cite{lee2020support,houston2018community}. A vital aspect of community organizing lies in its ability to create social support networks, which enhance improvisation and adaptability during recovery efforts \cite{lee2022patterns,lee2020support}. Such networks, however, can either hinder or accelerate resilience-building depending on their coordination and scope \cite{lee2020support}. This dynamic interplay highlights the role of communication in transitioning from collective action to systemic disaster responses. Grassroots organizations and hyperlocal networks are central to these collective recovery processes, often supplemented by emergent groups relying on informal communication channels to coordinate effective responses \cite{harris2022hyperlocal}. These efforts illustrate how bottom-up, multi-level communication—from neighborhood initiatives to broader inter-organizational landscapes—drives community resilience and recovery \cite{speer2012local,lee2020support,lee2022patterns,christens2015community}. 

\subsubsection{The effects of social media platform affordances on organizing possibilities}
\label{related_work_subsec:platform_posibilities}
Shklovski et al. argue that social media use increases in times of crisis as communities refocus on rebuilding and producing public goods \cite{shklovski2004internet}. However, people living in areas affected by crises are often overloaded with information from traditional sources (e.g., news) and online sources (e.g., social media) \cite[P.7]{soden_et_al_22}. The visibility of messages at the time of a crisis becomes a central concern to local advocates as they try to amplify messages about the resources. Twitter organizers used hashtags to search for relevant information and identify other organizers focusing on similar crises \cite{starbird_et_al_twitter_volunteer_2011}.  Another way to build an audience is by using mentions (e.g., @user). When a tweet mentions a user, other users can view and search for the tweet. The user can expand their audience by mentioning other accounts that have a bigger social network (e.g., a non-profit organization) \cite{mcgregor2019social}. Voting on the contents of a post is Reddit's filtering affordance. The downside of this affordance is that relevant posts might take too long to rise to the top. Multiple posts about the same topic in the same or in related subreddits means that multiple posts are competing for visibility. Moderators can create ``sticky'' posts that rise to the top regardless of the voting process to increase the salience of community emergencies \cite[P.1255]{Leavitt_robinson_17}.

Using Facebook pages, pet advocates attempting to broadcast messages about missing animals in Hurricane Sandy engaged in cross-posting - posting the same message to multiple Facebook pages. Because any Facebook user can post to Facebook pages (as opposed to Facebook groups where a poster needs to be a member), cross-posting allowed them to amplify messages about missing animals to different social networks \cite{white_et_al_cross_2014}. 

One of the affordances of Facebook is that social ties on the platform can mirror offline social ties. As outlined in \S\ref{subsubsec:comm_org}, these local ties become central to any local organizing effort since``citizen response is shaped both by primary social networks,  such as friends and family, and by altruistically motivated outreach to other victims nearby.'' \cite{birbak_FB_groups_snow_2012} The ability to tap into the equivalent of one's offline social networks in turn allows people to reconstruct their social network, or build community ties \cite[P.28]{semaan_mark_2012}. Indeed, because Facebook groups can be publicly searched, they can become an important organizing tool for local communities. Earlier work shows that the strong identity cues on Facebook (e.g., images showing people in their communities) creates more trust in local communication \cite{moser_et_al_17}. 

Kaewkitipong et al. \cite{KAEWKITIPONG2016653} studied the use of social media in the pre, during, and post-crisis phases. They found that social media plays important roles in each of the phases by ``fit[ting the information...knowledge[, and social support] needs of each phase.'' However, Facebook groups in rural areas have different dynamics and its members have different needs than those used in urban areas \cite{birbak_FB_groups_snow_2012}. Birkbak et al.\cite[P.434]{birbak_FB_groups_snow_2012} argue that online organizing in rural areas is more important to the local population due to geographical dispersion and lack of [traditional] media attention.'' This dynamic is reflected in Herdağdelen et al.'s work \cite{Herdağdelen_Adamic_State_2023} which identified latent factors that ``distinguished Facebook group engagement by county'' based on group locality, and group size, among other features. The authors found that rural Facebook groups tend to be public, medium sized (between 100 and 1,000 members), and age-diverse. They argue that this might be because in sparsely populated urban settings, most community members can participate in ``fairly big, well-mixed settings while [Facebook groups in more dense urban spaces]''...tend to be partially local (not hyper-local), bridging communities, large in membership size (more than 1,000 members) and usually associated with volunteering and local gift economies (i.e., donations) \cite[P.358-361]{herdaugdelen2023geography}.

%In a national-level analysis of Facebook groups, Herdağdelen et al. \cite{Herdağdelen_Adamic_State_2023} identified latent factors that ``distinguished Facebook group engagement by county'' based on group locality, and group size, among other features. Group locality measures the probability of any two randomly chosen members of a Facebook group being from the same county \cite[P.353]{Herdağdelen_Adamic_State_2023}. Group size is bucketed into four main groups: (1) very small (less than 30 members); (2) small (30<m<100); (3) mid-size (100< m <1,000); and (4) large groups (>1,000 members). One of the latent factors identified in Herdağdelen et al. fits our Facebook group. Specifically, they identify Facebook groups that are mid-sized or large, local, bridging members of community who might not have connections in offline social networks, especially in more densely populated, urban areas.    

\subsubsection{Local neighborhood groups organizing to cope with COVID}
\label{related_work_subsec:local_communities_COVID}
The enactment of social distancing guidelines during the COVID-19 pandemic led to an increased usage of social media \cite{cho2023bright}, thus opening up new spaces for people to meet locally, both online \cite{10.1145/3490632.3490666} and offline \cite{Haesler_et_al_stronger_together_21}. A surge in social media activity was driven not only by the availability of more time but also by the desire to stay socially connected and to access more information about COVID, and social support, especially in case the user is COVID-positive \cite{Jangid_et_al_together_alone_24}. Mutual aid groups were being formed online to organize local communities and support people facing food insecurity and other challenges associated with the COVID surge. Many of these mutual aid groups considered organizing to resolve long-standing root societal problems. This led to tension in those online mutual aid groups in what Soden et al. termed ``the long-now of community response to disaster.'' \cite[P.13]{soden_et_al_21}

However, in the early days of the pandemic, online mutual aid organizations focused on ``direct relief such as cash, groceries, or assistance with chores.'' \cite[P.9]{soden_et_al_21}. Immediate food aid, specifically for those who lost their jobs early in the pandemic, many of whom would have been stigmatized (or deemed ineligible) by asking for assistance from under-resourced social welfare organization, was especially important for mutual aid groups \cite[P.105]{Knearem_et_al_long_haul_21}. Other online support groups focused on supporting high-risk populations (e.g., senior citizens and immunocompromised populations) \cite[P.13]{Haesler_et_al_stronger_together_21}. Members of mutual aid groups managed Facebook pages and groups which served as information resources for local communities to better identify where community members can find the support they need \cite[P.236]{baca_muutal_aid_24}. Mutual aid organizations designed tools that meet the ``immediate needs of impacted communities.'' \cite{soden_et_al_21} like assistance hotlines and digital systems that track individual requests for assistance. 

As time progressed, these online communities started engaging in long-term food support by organizing "little free pantries and community fridges." \cite[P.105]{Knearem_et_al_long_haul_21} Other mutual aid groups supported social movements (e.g., black lives matter) by providing``physical support like [providing] water and snacks, or sharing information about protests on social media.'' \cite[P.11]{soden_et_al_21} Many of these groups started discussing organizing around a rent strike because of what they saw as the root problem of unaffordable housing. However, given the volunteer nature of these groups, such long-term organizing was seen as a challenge \cite{soden_et_al_21}.

\subsection{Public Sphere Theory}  \label{related_work_sec:public_sphere_theory}
This study turns to public sphere theory to understand how Online Neighborhood Group s and Town Hall Meetings can function as interconnected public spheres during disasters. By applying this theory, we examine how these spaces act as publics, allowing users to share information and organize around shared community issues.

\subsubsection{Publics} \label{related_work_subsec:publics}
Online Neighborhood Group s can give rise to \textit{publics} when members discuss and organize around shared issues \cite{maier2021talking}. The term public is multifaceted. As Van Dijk \cite{van1997study} explains, a public can be defined as (a) information awareness; or (b) an amorphous aggregate of individuals tied together \cite{van1985semantic}. Public as a quality of information involves the understanding that knowledge is widely known and shared, making ``collective thoughts and action possible'' \cite[P.14]{van1985semantic}. The shared awareness of information contributes to individuals' sense of collective power, or lack thereof, which can foster a sense of belonging or detachment \cite[P.14]{van1985semantic}. The second definition, public as an amorphous aggregate, is fundamental for self-governance in communities \cite[P.14]{van1985semantic}. When these two definitions are combined, the term public opinion emerges, meaning that making information public among an aggregate of people can lead to the publication of opinion. Public opinions are often shared in communicative spaces referred to as \textit{public spheres} \cite{habermas1989jurgen}.
In public spheres, people, audiences, or stakeholders discuss \textit{issues} (e.g., \cite{dewey1927half}), defined as ``socially constructed matters of public interest around which there is contention'' \cite[P.2009]{stoltenberg2024spaces}. Publics are not fixed or uniformed spaces \cite{friemel2023public}; they form and dissolve around issues (e.g., \cite{keinert2021relational}). In this way, issues are focal points around which publics coalesce \cite{bennett2014european, maier2018exploring}. As a result, publics can be divided horizontally into different groups (e.g., dominant publics, counterpublics; \cite{breese2011mapping, squires2005rethinking}) or vertically due to size (e.g., mass media, special interest; \cite{neuberger2022capture}), resulting in multiple publics within the same sphere \cite{neuberger2022capture}.

\subsubsection{Notions of space} \label{related_work_subsec:space_notions}
Maier et al. \cite[P.187]{maier2021talking} argues that public sphere theory ``still largely remains bound to territorial notions of space.'' Territorial notions of space include containers (e.g., rooms, homes) or surfaces (e.g., regions, states; \cite{waldherr2023translocal}, setting clear boundaries that delineate the inside from outside \cite{maier2021talking}. In HCI work, Semaan et al. \cite{Semaan2015public} has also called on sociotechnical theorists and researchers to embrace the sprawling nature of discourse in publics \cite{dahlgren2005internet} by not confining the public sphere to a single space. Online Neighborhood Group s, hosted on social media platforms like Nextdoor \cite{aubin2024not, choksi2024under}, have disrupted the territorial notions of public spheres as the ability to interact in these spaces are not bounded by physical or geographical limits \cite{hampton2016persistent}. Online Neighborhood Group s do not reside in any physical ``container'' but rather exist in a networked society (e.g., \cite{boyd2010social, castells2020information, habermas2006political}. Such studies challenge the foundation of public sphere theory, which traditionally recognized public discourse to manifest within well-defined physical boundaries, such as a town hall, a city square, or national media \cite{maier2021talking, waldherr2023translocal}. At the same time, studies on digital publics can ``leave one with the impression that such interaction takes place in a kind of virtual vacuum with little connection to the material worlds'' \cite[P.2]{jones2004problem}. The tendency to separate digital from material can be due to digital dualism \cite{jurgenson2011digital} or the assumption that such spaces operate under different communicative dynamics \cite{orgad2009can}. Yet prior work has described a symbiotic relationship between online and offline spaces \cite{albrechtslund2008online, lopez2015lend}.

\subsubsection{Online and offline spaces} \label{related_work_subsec:onlinevsoffline}
In community computing studies, online local systems are theorized ``to play a complementary role in the communication ecosystem of local communities'' \cite[P.66]{lopez2015lend}. As Willems \cite{willems2019politics} points out, scholars often treat digital and physical spaces as dualistic instead of interdependent or co-constitutive \cite[P.1194]{willems2019politics}. As a result, we are lacking empirical support for these assumptions as ``few studies have offered a detailed analysis of the intersection of, and continuity between, digital and physical spaces'' \cite[P.7]{willems2019politics}. The reason for this lack of studies can be due to offline data being ``difficult and expensive to collect'' \cite[P.227]{foucault2014dynamic}. Although limited, studies collecting perceptions have found some evidence to support the theoretical continuum between online and offline spaces in translocal communities. Erete \cite{erete2015engaging}, for example, interviewed residents that both engaged in local online discussions and attended community-police meetings. Despite the police department controlling the agenda, residents perceived the meeting to be community-driven due to the perceived similarity of online communication to in-person discussion \cite{erete2015engaging}. As Baym \cite[P.721]{baym2009call} argues, ``what happens via technology is completely interwoven with what happens face-to-face and via other media.'' 

Collectively, this work demonstrates that digital and offline communication can oftentimes form a symbiotic relationship \cite{ramirez2007online}.  For example, engaging in online activism has been correlated with future offline engagement advocating for a cause (e.g., \cite{10.1145/2470654.2470770}) especially when accessibility barriers limit one's ability to engage in offline activism (e.g., marching), having more fluid boundaries between online (e.g., e-petitioning) and hybrid (e.g., watching video streams) activism can make activism a more salient part one one's identity. This in turn would give people ``an internal motivation to continue performing and expanding their activist work.'' \cite[P.225]{li_slacktivists_2018} Following these calls for research, we aim to move away from comparing modalities in isolation and instead, examine how online communication functions in combination with face-to-face communication \cite{baym2009call, hall2020relating}.

\subsection{The materiality of physical and digital publics} \label{related_work_subsec:materiality_physical_digial}
The sociomateriality literature argues that users’ interactions with the material world are inseparable from their social practices \cite{barad2007meeting, suchman2008feminist}. Social and material worlds are co-constitutive, produced and enacted through one another \cite{orlikowski200810}. In essence, people perform their identities through a variety of sites, online and offline, through both material and discursive means. In this section, we start by introducing related work on materiality, followed by a discussion of materiality in traditional physical publics, and finally, comparing this to the materiality in digital publics.

\subsubsection{Materiality} \label{related_work_sec:materiality}Prior work emphasizes the importance of examining \textit{physical space} and \textit{materiality} (e.g., \cite{waldherr2021spaces, packer2013communication, couldry2004introduction, waldherr2023translocal, jansson2006towards, gillespie2014relevance}. Physical space refers to the geographically situated environments where communication and interaction take place \cite{waldherr2021spaces}. As Waldherr et al. \cite{waldherr2023translocal} argues, examining the physical space is important for understanding how the surroundings in which people communicate affect the nature and dynamics of those interactions. Materiality reflects the elements or artifacts present in these spaces, which can include technological affordances, furniture, architectural styles, and other material aspects that can facilitate or constrain interactions \cite{maier2021talking}. In HCI and CSCW studies, scholars investigating materiality examine how physical objects and their arrangements within a space can influence the flow and form of information and communication \cite{waldherr2023translocal}. For example, Semaan et al. \cite{Semaan2015public} stressed the importance of materiality of social media platforms in their study on individuals who used social media as a means for participation in the public sphere. The authors found that technological affordances that enabled customization helped users engage more effectively in civic activities \cite{Semaan2015public}. Therefore, we turn to prior work on the affordances of technological and physical settings to investigate the impact of such dimensions on community discourse.

Technological affordances are possibilities for action that represent “the relationship between individuals and their perceptions of environments” \cite[P.361]{parchoma_contested_2014}. Affordances of communication channels, such asynchronicity, the ability to communicate without constraints of time or space \cite{walther1996computer}, or bandwidth, the range of communication cues and media richness \cite{daft1986organizational}, can influence how communication processes unfold overtime \cite{evans_explicating_2017}.

\subsubsection{Materiality of traditional physical publics} \label{related_work_subsec:materiality_physical_traditional}
In his extensive ethnographic study, de Tocqueville \cite{de2017recollections} asserted that Town Hall Meetings are to liberty to what primary schools are to science; they teach individuals how to be responsible and participate in democracy. Such public forums are understood to be environments where free speech, inquiry, and community building thrive \cite{fennelly2011deliberative}. As Field \cite{field2019town} states, ``Generations of observers have celebrated the local, direct, and immediate nature of the town hall meeting as a particularly American form of democracy.'' Attending Town Hall Meetings can enhance feelings of intimacy and trust in organizational leaders by allowing citizens to interact directly with elected officials \cite{carr2018redressing}. These meetings are goal-oriented, which differs from the more open discussions found in online neighborhood forums \cite{hampton2016persistent}. Such forums are designed to be deliberative systems that use discourse as a means of problem solving \cite{townsend2022enduring}. The public discussion portion of Town Hall Meetings allow citizens to raise concerns that are not on the official agenda, providing individuals with an opportunity to express their grievances, opinions, and community-related news \cite{bryan2010real}.

Town Hall Meetings can have several limitations due to their physicality. For example, scholars often call into question the accessibility and representativeness of town hall forums. In-person meetings can exclude citizens with work obligations, lack of childcare, or those without transportation \cite{bryson2013designing, fung2015putting}. Neblo et al. \cite{neblo2010wants} found that attendees at Town Hall Meetings often have higher levels of education, income, and political interest, exacerbating the issue of representativeness. As a result, the views and concerns expressed at these meetings may not necessarily reflect those of the broader community \cite{gastil2005deliberative}. Field \cite{field2019town} argues that a lack of representativeness can limit the opportunities for genuine dialogue and compromise, which are essential for effective democratic deliberation. Prior work also argues that Town Hall Meetings can be subject to manipulation by bureaucratic organizations or special interest groups, compromising their ability to genuinely represent the public's interests \cite{field2019town, preertownhall}. Unequal power dynamics can arise when elected officials or influential community members dominate town hall discussions, which can discourage quieter or less confident attendees \cite{karpowitz2014deliberation} or those with dissenting viewpoints \cite{mansbridge2011clarifying} from participating.

\subsubsection{Materiality of digital publics} \label{related_work_materiality_}
Putnam \cite{putnam2000bowling} famously argued that technology is contributing to the decline in civic participation. While he has since recognized that social media can help promote civic conversations, Putnam \cite{putnam2020upswing} still posits that online forums are a mere starting point for engagement. Dotson \cite{dotson2017technically} takes a stronger stance, stating that communication technologies undermine in-person communication. Dotson \cite[P.7]{dotson2017technically} warns that a preference for online communication over face-to-face will make citizens ``poorly equipped to defeat better organized and funded elites in political conflict.'' However, prior work has found a positive correlation between engagement in virtual spaces and involvement in offline civic activities \cite{nah2022community, ognyanova2013online}. Such studies align with findings about the potential of neighborhood forums to cultivate social capital (i.e., resources embedded in social relationships; \cite{lin2017building}) and strengthen bonds among residents \cite{hampton2003neighboring, hampton2016persistent}. In addition to fostering connections, neighborhood forums can encourage community involvement by facilitating civic discussions about shared experiences \cite{hampton2017social}. Shared experiences can serve as a springboard for political discussion which can prepare ``citizens and the political system at large for political action'' \cite[P.20]{graham2014discursive}.

Neighborhoods can exemplify Anderson’s \cite{anderson2020imagined} concept of \textit{imagined communities}, defined as social groups whose members might never meet face-to-face but maintain a sense of connectedness. As Anderson \cite[P.6]{anderson2020imagined} explains, “members of even the smallest nation will never know most of their fellow-members…yet in the minds of each lives the image of their communion.” The concept of imagined communities is often applied to Internet studies to describe how online users build a sense of connectedness with others despite not knowing each other \cite{gruzd2011imagining}. The distinction between what is considered public and private can become more fluid in digital spaces \cite{papacharissi2010private}, leading to a context collapse \cite{marwick2011tweet}. For example, public discussions in Online Neighborhood Group s are accessible from private spaces (e.g., one’s home), and more private conversations can be public when shared through shared on social networking sites. This blurring of distinction between private and public can alter ``the actual and imagined spaces upon which citizenship is practiced'' \cite[P.17]{papacharissi2010private}. As boyd \cite[P.49]{boyd2010social} explained, ``the lack of spatial, social, and temporal boundaries makes it difficult to maintain distinct social contexts.'' However, scholars have argued that in-person spaces offer physical co-presence \cite{short1976theoretical}, defined as being in the same physical space as others, allowing for face-to-face communication \cite{rice1992task}.

Spaces that afford physical co-presence often involve non-verbal cues, immediate feedback, and similar sensory experiences that are difficult to replicate in mediated communication \cite{tian2021physical}. Still, HCI scholars have found that users can still feel connected to others online despite a lack of physical co-presence \cite{randazzotrauma, ammarisaod}. Goffman \cite{goffman1959presentation} uses the concepts of ‘front stage’ and ‘back stage’ to theorize the reasons as to why social connection can still be achieved online despite the lack of face-to-face interaction. front stage refers to the part of one’s identity that one performs and enact roles for others \cite{goffman1959presentation}. back stage is where individuals can express themselves more freely and drop their social mask \cite{goffman1959presentation}. Scholars using this metaphor suggest that users feel more comfortable being themselves online as opposed to in-person settings. Hampton \cite{hampton2017social} argues that the asynchronous nature of neighborhood forums reduces the pressure to formulate immediate responses. While it is true that users can contribute to discussions at their own convenience, prior work suggests that this flexibility can be overridden when there is significant pressure from pressing community concerns. During disruptive events, the need for citizens and organizations to organize collective action \cite{doerfel2017story}, defined as two or more individuals in pursuit of a shared benefit \cite{olson1965logic}. Disruptive events can create a sense of urgency that permeates online forums (\cite{dahlberg2016tech, gerbaudo2012tweets, gonzalez2016networked}.

Collectively, the above literature allows us to investigate disaster discourse during the short and long-term recovery phases of a disaster and if and how online and offline publics co-constitute each other.
\begin{quote}
    \textbf{RQ1:} What are the discourse themes in an Online Neighborhood Group ? 
\end{quote}
\begin{quote}
    \textbf{RQ2:}  What are the discourse themes in community Town Hall Meetings? 
\end{quote}
\begin{quote}
    \textbf{RQ3:} How similar or dissimilar are topics across online neighborhood threads and offline Town Hall Meetings?
\end{quote}

\section{STUDY DESIGN}
In this section, we start by describing our data collection process for both the online community as well as the Town Hall Meetings. We then describe our methods in detail. 

\subsection{Data Collection} \label{sec:methods_data_collection}
After receiving Institutional Review Board (IRB) approval, we chose a densely populated community (n = 12,000 residents) in the New York Metropolitan region. Around 46.8\% of the borough’s most at-risk residents experience three or more resilience risk factors (e.g., income-to-poverty ratio, overcrowding, households lacking full-time, year-round employment) which is higher than the national average \cite{censusresilience}. These risk factors played a significant part in our decision to select this location. 

For Corpus 1, the town hall dataset, we collected the public discussion portion of transcripts from YouTube recordings of Town Hall Meetings. To validate the accuracy of the transcripts, a research assistant listened to the original recordings and manually edited errors, generating 67 single-spaced pages of combined transcriptions. Throughout the period of the study, there were 51 unique town hall meeting contributors. The board of county commissioners sets the agenda for the Town Hall Meetings, and it was not possible for community members to propose topics. 

For Corpus 2, the online dataset, we used CrowdTangle's application programming interface (API; \cite{crowdtangle}) to retrieve posts from the Online Neighborhood Group  on Facebook. The Facebook group, created in 2014, is a general neighborhood group, not created for the hurricane. There were a total of 6,284 members in the Facebook Group and 2,463 of them were active contributors at the time of data collection. Conversation threads on social media \cite{shugars2019structure} are often used to examine community discourse \cite{aubin2024not, xin2019assessing}. We collected 15,021 observations including posts and associated comments covering a period of eight months between September 1, 2021, when Hurricane Ida first made landfall in the Northeast and May 16, 2022. Given our interest in comparing online vs. offline discourse, we treated these data as cross-sectional. 

Twenty-five unique community members are active members of the Facebook group who also contributed to Town Hall Meetings.

\subsection{Data Analysis} \label{sec:methods_data_analysis}
\begin{figure*}[ht]
  \centering
  \includegraphics[width=.8\textwidth]{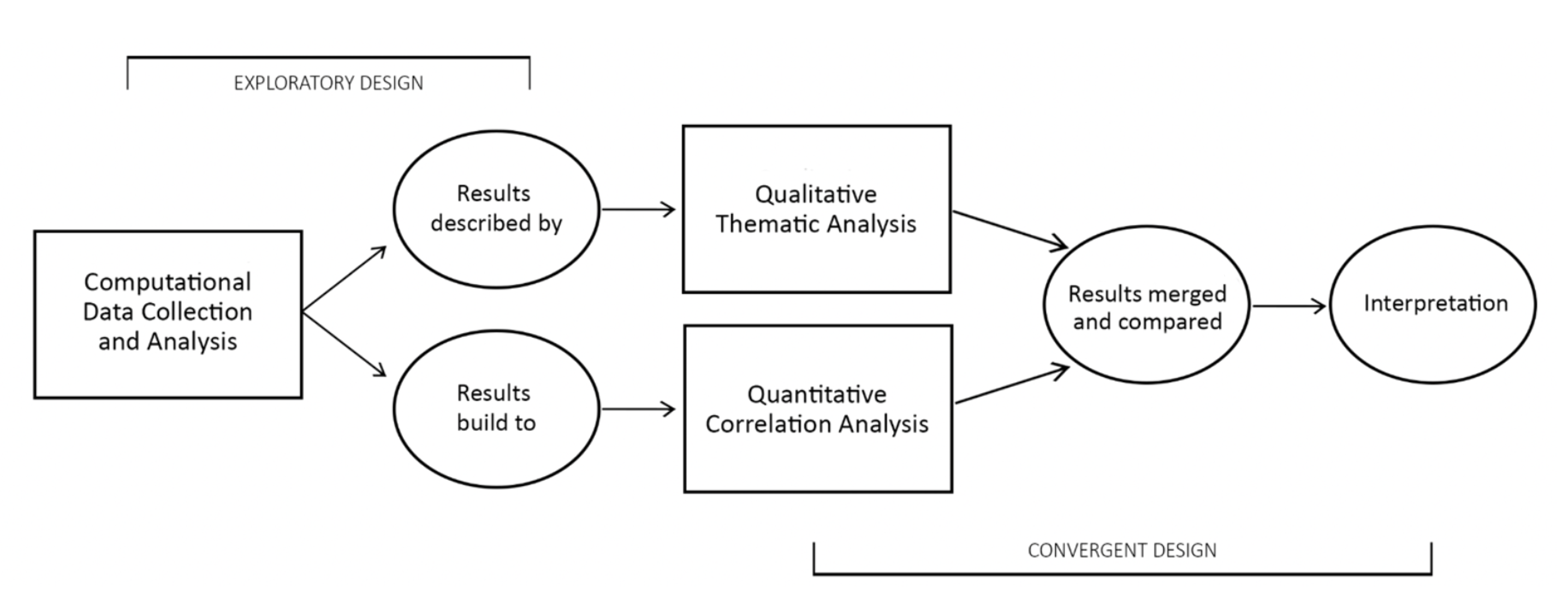}
  \caption{Study Design: after collecting data from both Town Hall Meetings and Online Neighborhood Group  threads on Facebook, we used computational methods to describe the topics discussed in both. We then conducted thematic analysis to contextualize and thematically group the topics.}
  \Description{This study uses a mixed method convergent design: after collecting data from both Town Hall Meetings and Online Neighborhood Group  threads on Facebook, we used computational methods to describe the topics discussed in both. We then conducted thematic analysis to contextualize and thematically group the topics.}
  \label{fig:studydesign}
\end{figure*}

This study uses Creswell and Clark's \cite{creswell2017designing} convergent parallel design \cite{creswell2017designing}. This approach  involves conducting quantitative and qualitative data collection and analysis concurrently. The purpose of this design is to capitalize on the strengths of both quantitative and qualitative methods, allowing for an in-depth exploration of the research problem from multiple angles. We executed this design using the steps depicted in Figure \ref{fig:studydesign} which we describe in more detail in the below section. Ultimately, the convergent parallel design helped facilitate a deeper understanding and more comprehensive interpretation of community discourse in online and offline settings.

\subsubsection{Topic modeling} \label{subsec:method_topic_modeling}
We utilized topic modeling, an unsupervised natural language processing tool (NLP) that allows researchers to uncover topics in large corpora of text \cite{choi2020digital, gad2015theme}. We turned to latent Dirichlet allocation (LDA), one of the most popular algorithms for topic modeling \cite{paulus2019looking} and Gensim’s LDA implementation for analysis \cite{rehurek2020gensim}. LDA identifies a set of underlying topics by examining patterns of co-occurring words within collections of documents \cite{chang2010hierarchical}. First, we pre-processed both datasets, which included removing punctuation, stop words, symbols, and numbers, and conducting lemmatization (i.e., reducing different forms of a word to one single form; \cite{doerfel1998constitutes}).
For Corpus 1 (town hall), documents represented each instance a resident spoke during the public discussion portion of Town Hall Meetings.

\begin{figure*}[ht]
  \centering
  \includegraphics[width=.8\textwidth]{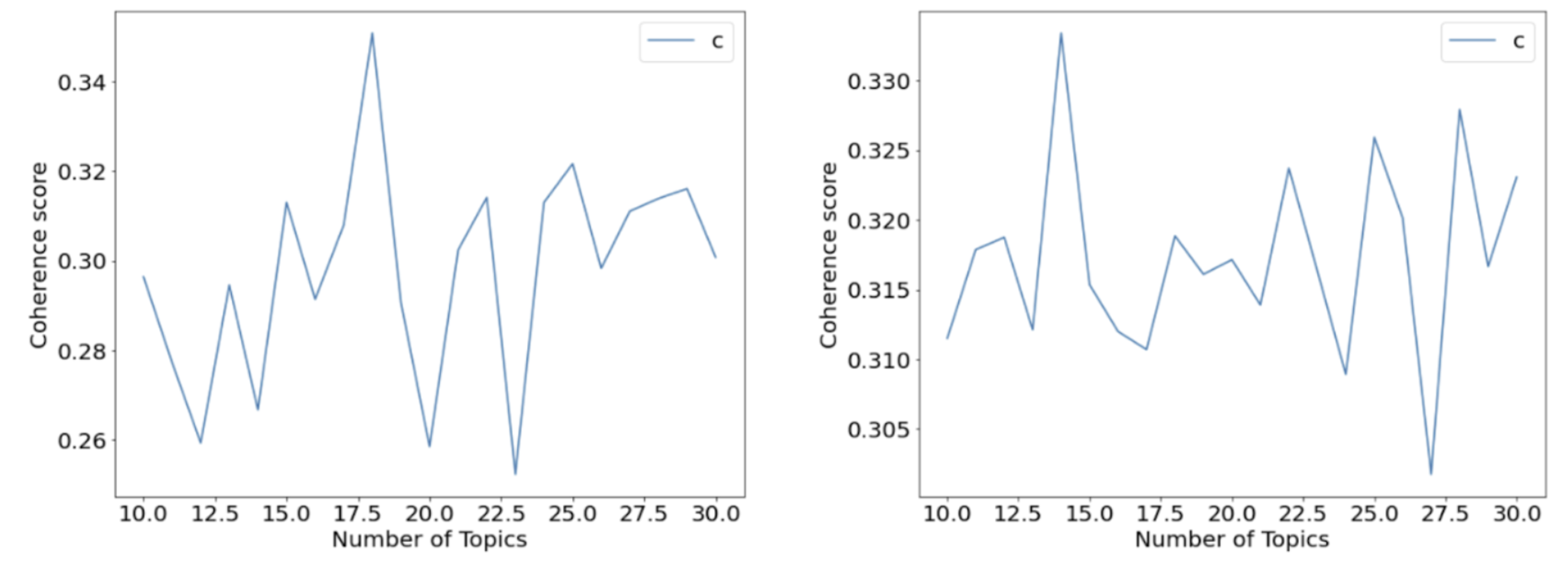}
  \caption{Coherence Scores of Topic Models. The left graph shows the coherence scores for Town Hall Meetings which peak at 18 topics. The right graph shows the coherence scores for the Online Neighborhood Group  which peak at 14 topics.}
  \Description{The coherence scores peak at k = 18 for the Town Hall Meetings and k = 14 topics for the online threads. }
  \label{fig:coherence}
\end{figure*}

For Corpus 2 (online), we grouped Facebook posts and corresponding comments into separate documents to capture the topics discussed within a single thread, considering each thread to be a document in the online corpus. We trained 30 LDA models for each corpus to determine the optimal number of topics relying on coherence scores (using Gensim's \textit{CoherenceModel} \footnote{\url{https://radimrehurek.com/gensim/models/coherencemodel.html}} feature) which have been found to be better at approximating human ratings of a topic model ``understandability'' \cite{roder_exploring_2015,chang_reading_2009}. The coherence scores peak at k = 18 for the Town Hall Meetings and k = 14 topics for the online threads (Fig. \ref{fig:coherence}). Three members of the research team compared the keywords of the LDA models with similar coherence scores and confirmed that the peak coherence scores were the most interpretable models.

\subsubsection{Jaccard Similarity} \label{subsec:method_jaccard}
\begin{figure*}[ht]
  \centering
  \includegraphics[width=.8\textwidth]{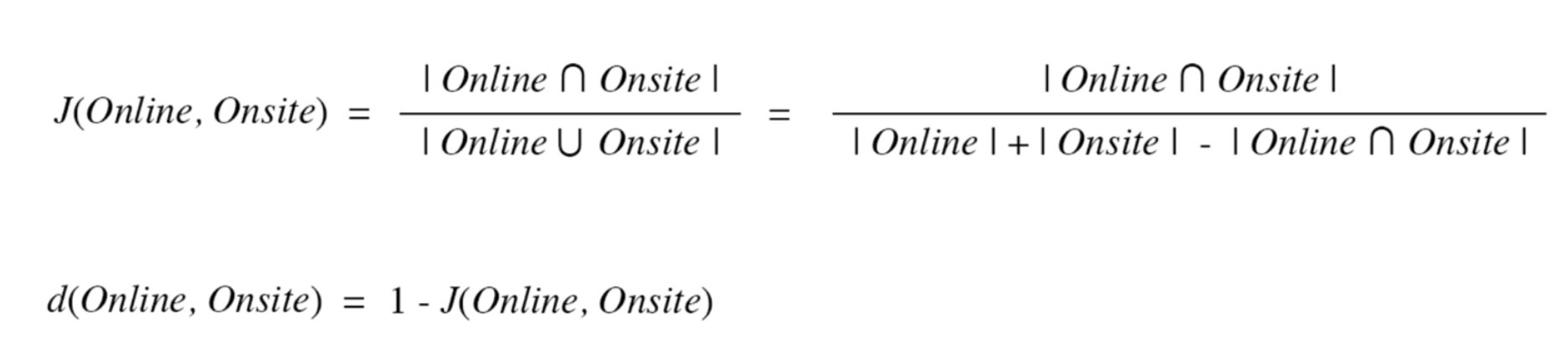}
  \caption{Jaccard Coefficients}
  \Description{Equations of the Jaccard similarity and dissimilarity coefficients}
  \label{fig:jaccard}
\end{figure*}

We used Jaccard similarity coefficients to measure the similarity between Town Hall Meetings and the Online Neighborhood Group datasets \cite{jaccard1912distribution}. When applied to topic modeling, the Jaccard index measures how similar two topics are based on their words or documents \cite{ammari2018jaccard}. For example, two topics that both contain a significant number of the same terms will have a higher Jaccard score, indicating a strong similarity \cite{li2019baby}. This method is particularly helpful when analyzing the overlap or distinction between thematic content of discussions across different communication platforms \cite{rehurek2020gensim}. Figure \ref{fig:jaccard} shows the formula used to calculate the Jaccard scores \cite{rehurek2020gensim}. The Jaccard coefficient measures J (the coefficient) by dividing the size of the intersection in both datasets ($A \cap B$) by the number of unions in either set ($A \cup B$) and measures d (distance) by calculating $1-J(A,B)$ \cite{karabiberjaccard}. The Jaccard coefficient for topics across the corpora is represented as a heatmap in Figure \ref{fig:heatmap}. Some of the topics were highly similar (\verb|~|0.99), while others registered more moderate similarity scores (\verb|~|0.7).

\subsubsection{Thematic analysis} \label{subsec:methods_thematic_analysis}
We applied thematic analysis to contextualize and thematically group topics \cite{corbin2014basics}. To manage the scope of the analysis, we employed a random sampling technique to compile observations (e.g., posts, comments, transcripts) for each identified topic. Due to the length of observations from Town Hall Meetings, the random samples consisted of two documents per topic, each approximately a half page of single-spaced text (\verb|~|550 words). Posts from the online corpus were significantly shorter (\verb|~|66 words each). As a result, we sampled 10 observations per topic to qualitatively analyze the online corpora. Following Braun and Clarke’s \cite{braun2006using} guidelines for qualitative coding, three members of the research team independently reviewed the samples and identified preliminary themes. During this initial phase, we utilized open coding techniques which allowed for new insights to emerge without the constraints of preconceived categories \cite{braun2006using}.

We then discussed and compared our findings which helped us develop a coding scheme. This scheme included both thematic categories (e.g., housing concerns, public health) and subcategories (e.g., COVID-19 testing kits, opposition to rezoning). In the next phase, we conducted axial coding which involved examining the relationships between codes. An example of axial coding includes asking the following question, how does ``COVID-19'' and ``opposition to rezoning'' overlap or influence each other, if at all? In our team meetings, we meticulously discussed different thematic bridges between codes during the axial coding phase and discrepancies in our interpretations. We defined the core categories during the selective coding phase. During this final step, the entire team decided which themes are most central to the understanding of each individual topic. As Braun and Clarke \cite{braun2006using} explain, selective coding aims to ensure that each theme is rigorously discussed and the coding scheme adjusted based on the team’s collective insights. This process was also central in determining distinctions between discourse in Townhall meetings and Online Neighborhood Group  discussions showed a high similarity.

\paragraph{A note on quoting Facebook posts} Fiesler and Proferes \cite{fiesler_participant_2018} showed that social media users do not expect to be quoted verbatim in academic research. We followed  Bruckman’s \cite{bruckman2002studying} recommendations which involved describing user quotes without quoting them directly and changing minor details to protect the privacy of group members. Similar strategies for quoting online community users, especially sensitive discussions, were used in earlier literature (e.g., \cite{ammari_throw_19}). 

\section{FINDINGS}
In this section, we first present our answers to RQ1 by describing the discourse in our online environment. Next, we analyze the discourse in the offline environments to answer RQ2. Finally, the answer to RQ3 is then presented by showing the similarity and dissimilarity between the two. 

\subsection{RQ1: What are the discourse themes in an Online Neighborhood Group ?}

As seen in Table \ref{tab:topic_models}, 14 topics emerged from the LDA model for the online corpus. Using thematic analysis, we organized the topics into broad themes (e.g., social support, recovery) to give insight into the online topics following a natural disaster. The topics are labeled as the following: Topic ONG \textit{Mobilizing social capital}, Topic ONG2 \textit{Local knowledge}, Topic ONG3 \textit{COVID-19}, Topic ONG4 \textit{Educational resources}, Topic ONG5 \textit{Policy and governance}, Topic ONG6 \textit{Well-being updates}, Topic ONG7 \textit{Advice}, Topic ONG8 \textit{Developer criticisms and solutions}, Topic ONG9 \textit{Local activities}, Topic ONG10 \textit{Life-or-death situations}, Topic ONG11 \textit{Emergency management and assistance}, Topic ONG12 \textit{Seeking recommendations}, Topic ONG13 \textit{Announcements and news}, and Topic ONG14 \textit{Donations}.

\subsubsection{Social support} \label{subsec:results_social_sup} The topics reflected an exchange of informational , emotional support, and monetary support. For instance, Topic ONG2 (local knowledge) represented community members sharing specialized or local knowledge, mainly about living conditions for local renters. We observed users explaining that the poor treatment of residents, such as not providing working heat, is nothing new. Other comments disclosed their negative experiences renting and shared recommendations on how to organize and exercise their rights as renters. Topic ONG3 (COVID-19) was about users' experiences waiting in long lines to obtain booster shots or coming across empty shelves when seeking testing kits. At the time, the United States was experiencing a national coronavirus test shortage \cite{pietschcorona}. Online users attempted to circumvent the shortage by informing each other of which stores had testing kits available. The comments for Topic ONG3 also revealed concerns about the lack of enforcement of mask mandates in K-12 schools.

Topic ONG4 (educational resources) contained comments about educational resources (e.g., guides, tutoring services) while Topic ONG12 (seeking recommendations) included users looking for referrals for local service providers (e.g., contractors, lawyers). Topic ONG7 (advice) was personal while still focused on information sharing. For example, one user said, ``After [removed] died in the flash flooding, her family is urging others to purchase a safety hammer to prevent a similar tragedy from happening.'' Topic ONG10 (life-or-death situations) exemplified emotional support and sensemaking. Users discussed local tragedies (e.g., house explosion, COVID-19) that resulted in the deaths of community members. These discussions mostly encouraged concern and empathy for each other, using language such as ``I'm sorry to hear'' or ``deepest condolences.'' However, some comments can be perceived as controversial by hinting at conspiracy theories (e.g., overcounting of COVID-19 deaths; \cite{aschwanden2020}).

\subsubsection{Recovery} \label{subsec:results_recovery} Topics about recovering from the storm were categorized as either long-term or short-term recovery. Short-term recovery included topics that focus on immediate needs such as clothing, shelter, and food. Topics under long-term recovery pertained to long-term needs such as access to safe and affordable housing. Topic ONG1 (mobilizing social capital) pointed to the immense self-organizing citizens often do after an environmental disaster. In this sample, comments mentioned partnerships with community-based organizations (e.g., first responders, churches) and requests for volunteers to support cleanup and donation efforts. Topic ONG11 (emergency management and assistance) contained comments that relayed information from the Federal Management and Emergency Agency (FEMA), a government department that helps United States citizens mitigate risks (FEMA, 2013). For example, one comment said, ``This is the last day to talk to FEMA in person. They are at [location] across...If you know people who do not have access to this information online, please call and share this message with them.'' As this comment demonstrates, topics focused on recovery highlight the importance of information dissemination in post-disaster contexts.

Topic ONG14 (donations) was mainly about donations beyond the essentials (e.g., shelter, food) such as children's toys that were destroyed during the storm. One comment said, ``I love that people are helping locate toys and not just donating water and meals for families. Things like this helps to ease trauma for the kids.'' Long-term recovery pointed to the lack of policies protecting low-income residents. For instance, one comment in Topic ONG5 (policy and governance) said, ``The push against affordable housing has been going on for decades...the arrogance of the mayor at the last council meeting makes me feel sick.'' Comments in Topic ONG5 included discussions of blame assignment which often happens when citizens perceive responsible actors as unwilling to help or when standard explanations for a tragedy have failed. For example, one user commented,
\begin{quote}
There is one post claiming that the council was completely surprised by the storm. Another saying that the fault lies in the previous administration. Another screaming about Houston. Others giving the climate change argument a shot. Everyone is deflecting blame. The town has done great things, but preparation for the storm was not one of them.    
\end{quote}
This quote reflects the frustrations and challenges related to long-term recovery and policy shortcomings. Users' views on affordable housing policies and the perceived lack of preparation for the storm reveal deep-seated issues in governance and disaster management. 

Long-term recovery discussions also encompassed Topic ONG8 (Developer Criticisms and Solutions), where the conversation focuses on a real estate developer interested in demolishing low-income housing affected by Hurricane Ida. One comment in the group stated, ``Pretty soon [developer] will want their name up on the buildings in town. Maybe even a street named after them. Slumlord.'' Referring to the developer as a ``slumlord'' exemplified the platform's role in fostering unfiltered public expression and highlighted the emotional and contentious nature of the discussions. The group also served as a space for proposing actionable steps, such as filing ``a class action lawsuit'' or joining a ``free online support group.'' These suggestions are aimed at providing both legal recourse for those affected by the disaster. This dynamic showcased the group's function as a place for sharing frustrations and for community members to organize and seek collective solutions to the challenges they faced following the disaster. 

\subsubsection{Local news} The Online Neighborhood Group  was an active platform for local news (Topics ONG6, ONG9, \& ONG13). Topic ONG6 (well-being updates), Topic ONG9 (local activities), and Topic ONG13 (announcements and news) were reminiscent of information often stapled to telephone poles or pinned to community bulletin boards. These topics included posts that shared announcements and local news, updates well-being (e.g., finding car keys), and local activities (e.g., festivals, holiday celebrations, parades). For example, one user commented, ``In April, there will be vendors and food trucks. Meet our horses, pigs, and chickens. Parking fee goes to animal care.'' However, unlike community bulletin boards, Online Neighborhood Group s afford connectivity which allowed for neighbors to update each other on their well-being. For example, in Topic ONG6, one user wrote, ``Update. My keys have been found. Thank you to everyone in the group that tried to help. And a sincere thanks to [group member] for finding my keys!'' While not pertinent to wide-spread community issues, online discussions around mundane or everyday topics, like loosing one's keys, can still serve a function by helping individuals bridge connections and build social capital. Collectively, these results demonstrate how the Online Neighborhood Group  functioned as an open, informal space, often resembling a community bulletin board.

% \usepackage{multirow}
\begin{table}[]
\begin{tabular}{|l|l|c|l|l|}
\hline
\multicolumn{2}{|c|}{\textbf{Town Hall}}                                                  & \multirow{2}{*}{} & \multicolumn{2}{c|}{\textbf{Online Neighborhood Group }}                                        \\ \cline{1-2} \cline{4-5} 
\multicolumn{1}{|c|}{\textit{\textbf{No.}}} & \multicolumn{1}{c|}{\textit{\textbf{Topic Name}}} &                    & \multicolumn{1}{c|}{\textit{\textbf{No.}}} & \multicolumn{1}{c|}{\textit{\textbf{Topic Name}}} \\ \cline{1-2} \cline{4-5} 
\multicolumn{1}{|l|}{TH1}                   & Town Appreciation                                 &                    & \multicolumn{1}{l|}{ONG1}                  & Mobilizing Social Capital                         \\ \cline{1-2} \cline{4-5} 
\multicolumn{1}{|l|}{TH2}                   & Homeowner Concerns                                &                    & \multicolumn{1}{l|}{ONG2}                  & Local Knowledge                                   \\ \cline{1-2} \cline{4-5} 
\multicolumn{1}{|l|}{TH3}                   & Flood Mitigation                                  &                    & \multicolumn{1}{l|}{ONG3}                  & COVID-19                                          \\ \cline{1-2} \cline{4-5} 
\multicolumn{1}{|l|}{TH4}                   & Unsafe Living Conditions                          &                    & \multicolumn{1}{l|}{ONG4}                  & Educational Resources                             \\ \cline{1-2} \cline{4-5} 
\multicolumn{1}{|l|}{TH5}                   & Urging Change                                     &                    & \multicolumn{1}{l|}{ONG5}                  & Policy \& Governance                              \\ \cline{1-2} \cline{4-5} 
\multicolumn{1}{|l|}{TH6}                   & Condemning \& Defending Developer                  &                    & \multicolumn{1}{l|}{ONG6}                  & Wellbeing Update                                  \\ \cline{1-2} \cline{4-5} 
\multicolumn{1}{|l|}{TH7}                   & Overdevelopment                                   &                    & \multicolumn{1}{l|}{ONG7}                  & Advice                                            \\ \cline{1-2} \cline{4-5} 
\multicolumn{1}{|l|}{TH8}                   & Community-based Organizations                     &                    & \multicolumn{1}{l|}{ONG8}                  & Developer Criticisms \& Solutions                 \\ \cline{1-2} \cline{4-5} 
\multicolumn{1}{|l|}{TH9}                   & Future Impact of Construction                     &                    & \multicolumn{1}{l|}{ONG9}                  & Local Activities                                  \\ \cline{1-2} \cline{4-5} 
\multicolumn{1}{|l|}{TH10}                  & Government Negligence                             &                    & \multicolumn{1}{l|}{ONG10}                 & Life-or-Death Situations                          \\ \cline{1-2} \cline{4-5} 
\multicolumn{1}{|l|}{TH11}                  & Public Works                                      &                    & \multicolumn{1}{l|}{ONG11}                 & Emergency Management                              \\ \cline{1-2} \cline{4-5} 
\multicolumn{1}{|l|}{TH12}                  & Overcrowding                                      &                    & \multicolumn{1}{l|}{ONG12}                 & Seeking Recommendations                           \\ \cline{1-2} \cline{4-5} 
\multicolumn{1}{|l|}{TH13}                  & Storm Damages                                     &                    & \multicolumn{1}{l|}{ONG13}                 & Announcements \& News                             \\ \cline{1-2} \cline{4-5} 
\multicolumn{1}{|l|}{TH14}                  & Infrastructure Breakdown                          &                    & \multicolumn{1}{l|}{ONG14}                 & Donations                                         \\ \cline{1-2} \cline{4-5} 
\multicolumn{1}{|l|}{TH15}                  & History of Developer                              &                    & \multicolumn{1}{l|}{}                      &                                                   \\ \cline{1-2} \cline{4-5} 
\multicolumn{1}{|l|}{TH16}                  & Affordable Housing                                &                    & \multicolumn{1}{l|}{}                      &                                                   \\ \cline{1-2} \cline{4-5} 
\multicolumn{1}{|l|}{TH17}                  & Citizen Advisory Boards                           &                    & \multicolumn{1}{l|}{}                      &                                                   \\ \cline{1-2} \cline{4-5} 
\multicolumn{1}{|l|}{TH18}                  & Budgets                                           &                    & \multicolumn{1}{l|}{}                      &                                                   \\ \cline{1-2} \cline{4-5} 
\end{tabular}
\caption{Table shows Town Hall meeting topics and Online Neighborhood Group  topics.}
\label{tab:topic_models}
\end{table}

\subsection{RQ2: What are the discourse themes in community Town Hall Meetings?} 
A total of 18 topics surfaced from the LDA model for the town hall corpus. We organized these topics into high-level themes (e.g., infrastructure needs, community needs, citizen feedback) to investigate the public discussion portion of Town Hall Meetings. The 18 topics are labeled as the following: Topic TH1 \textit{Town appreciation}, Topic TH2 \textit{Homeowner concerns}, Topic TH3 \textit{Flood mitigation}, Topic TH4 \textit{Unsafe living conditions}, Topic TH5 \textit{Raising awareness and urging change}, Topic TH6 \textit{Condemning and defending developer}, Topic TH7 \textit{Overdevelopment}, Topic TH8 \textit{Community-based organizations}, Topic TH9 \textit{Future impact of construction}, Topic TH10 \textit{Government negligence}, Topic TH11 \textit{Public works}, Topic TH12 \textit{Overcrowding}, Topic TH13 \textit{Storm damages}, Topic TH14 \textit{Infrastructure breakdown}, Topic TH15 \textit{History of developer}, Topic TH16 \textit{Affordable housing}, Topic TH17 \textit{Citizen advisory boards}, and Topic TH18 \textit{Budgets}.

\subsubsection{Infrastructure needs.} Topics under this category concentrated on the management and maintenance of facilities. In Topic TH3 (flood mitigation), the public addressed the council for not mitigating property damages by moving the town's emergency vehicles to a higher elevation. For example, a community member said, ``There was tremendous property loss that I think could have been mitigated. I would like to see an action report that says how we can do better.'' Topic TH7 (overdevelopment) shed light on the public's lack of confidence in the borough's infrastructure to withstand future environmental disasters. ``They were talking about this in the resolution, about impervious coverage...that water has got to go someplace.''\footnote{Impervious coverage is a human-made surface that resists rainfall such as rooftops, patios, and walkways \cite{tropcyclone2020}}. For some citizens, an increase in development means less nature to absorb excess rainwater. Similarly, Topic TH12 (overcrowding) pointed out the consequences of high-density housing (i.e., developments with a higher population than average). For example, a member of the public said,
\begin{quote}
    Rezoning to create more density, in my opinion, is a mistake. We've created tremendous density in this town over the last few years with some wonderful new developments coming in but I feel that we've fulfilled our responsibility. There is actually a very dangerous pattern here…You know that we've got flood zones in the area. We don't need to build structures in flood zones.
\end{quote}
This quote highlighted a critical need for reform in the town's development policies and emergency preparedness plans. This theme also included Topic TH14 (infrastructure breakdown), which was composed of public comments that questioned the town's decision to sell donated emergency vehicles (e.g., rescue boat, reserve fire engine) before the storm. The citizen recommended the town council ``have a consultant come in and look at our organizational structure'' to avoid ``major bad decisions being made in this town.''

\subsubsection{Community Needs.} Findings revealed that residents used the town hall forum to raise awareness of community needs. Topic TH4 (unsafe living conditions), Topic TH17 (citizen advisory boards), and Topic TH18 (budgets) all highlighted the need for inter-organizational collaboration between the residents, the town council, and local nonprofit organizations. Inter-organizational partnerships are essential for solving complex problems, especially after a disaster. For instance, a resident from one of the town’s low-income apartments said,
\begin{quote}
    The [apartment complex] is now requiring non-disclosure agreements or NDAs in order for displaced residents to either break their lease, get their security deposits, or even get their rent payment refund for the time they've been displaced…Vulnerable residents in our community are being forced to silence their voice. These concerns affect [town's name] as a whole. If tax money for affordable housing is going to subpar if not dangerous accommodations then it is in everyone's best interest to be aware of those issues. Properties requiring NDAs makes it impossible to make issues like this public…We are asking you, the borough council, to step in and help us resolve these issues. How can we even attempt to recover from the devastating effects of Hurricane Ida if [apartment complex] actively refuses to speak with us?
\end{quote}

This quote exemplified the community's need for collaboration with the town council to enact change. Topic TH17 (citizen advisory boards) highlighted the community's call for collaboration through citizen advisory committees. These committees allowed residents to actively participate in assessing local needs and crafting solutions over time. In Topic TH8, comments often came from individuals affiliated with community organizations such as the women’s civic league, which was praised for its courage and support during and after the hurricane. Such community-based organizations, including nonprofits and churches, played a crucial role in resilience by effectively mobilizing volunteers and managing donations. Topic TH16 (affordable housing) revealed how climate change subtly influenced everyday life and the interconnected needs of the community, such as economic development and accessible housing. In one comment, a resident said
\begin{quote}
    I know that back in the 90s, the challenge was the mall and how it decimated downtown. There's been a lot of changes since then, most of them good but the challenge now is managing growth and managing that growth in the context of climate change. I encourage the board to make decisions that make [town name] resilient to climate change and affordable to residents.
\end{quote}    
Not all public comments offered solutions. Most residents used the town hall forum to share their concerns and asked for the governing body to address these issues. In Topic TH5 (urging change), a long-time resident said, 
\begin{quote}
    I live in the most ethnically and economically diverse portion of this town. There are people in my neighborhood who need every dime and dollar that they can get from the federal government, the state government, section 8, which is a program that I’ve wanted in this town a long time ago. We have people with six figure incomes right around the corner. We’re going to lose that...We’re losing our neighborhoods. We’re losing our community.
\end{quote}
Similarly, in Topic TH2 (homeowner concerns), residents shared concerns related to owning a home in the borough (e.g., mail mix-ups, changing neighborhood). One resident said,
\begin{quote}
Any buildings taller than three stories in a residential area is going to get rid of that neighborhood feel that we have. And I really implore you to consider that, to keep the downtown a downtown and keep our neighborhood feeling and looking like a real neighborhood.    
\end{quote}
This plea reflected the broader community's desire to maintain a balance between development and the preservation of local charm. Residents' concerns emphasized the importance of thoughtful urban planning that respects the cultural aspects of the neighborhood in addition to its aesthetic. Engaging with such feedback is crucial for city officials and planners in ensuring that development projects enhance rather than detract from the community's existing character and livability. Collectively, this theme illustrated the neighborhood's strong desire for a collaborative relationship with the town council to address pressing community issues.

\subsubsection{Citizen Feedback.} This theme included feedback that ranged from complimentary such as Topic TH1 (town appreciation) to critical like Topic TH10 (government negligence). While some residents expressed gratitude, others were disappointed with the town’s emergency response (or lack thereof) to Hurricane Ida and actions of local landlords. One resident in Topic TH10 explained how elected officials failed to learn from past weather events, ``We are already aware there was a gross lack of preparedness prior to the storm. To not look back and prepare like we have in the past when we've dealt with flooding…is negligence. So now, as always, the community came together in the aftermath.'' In this quote, the accusation of negligence regarding disaster preparedness underscored the residents' call for accountability. The quote also highlights the residents' expectations that lessons should be learned from past experiences with similar crises.

Topic TH6 (condemning and defending the developer) contained a mix of positive and negative feedback on the real estate developer whose company managed both luxury and low-income properties across the borough. In defense of the developer, one attendee said, ``I am an attorney with [removed] law firm. I represent [apartment complex]...we have members of the fire department, code officials strategizing trying to find the most effective path to getting residents back to their dwellings...I believe my client has done as much as it possibly could.'' The contrasting views within this topic reflect the complexity of public opinion regarding urban development and set the stage for a deeper exploration in RQ2. Overall, the diversity of opinions presented reflected the community's engagement and interest in the governance and resilience of their neighborhood.

\subsection{RQ3: How similar or dissimilar are topics across online neighborhood threads and offline Town Hall Meetings?}
\begin{figure*}[ht]
  \centering
  \includegraphics[width=.8\textwidth]{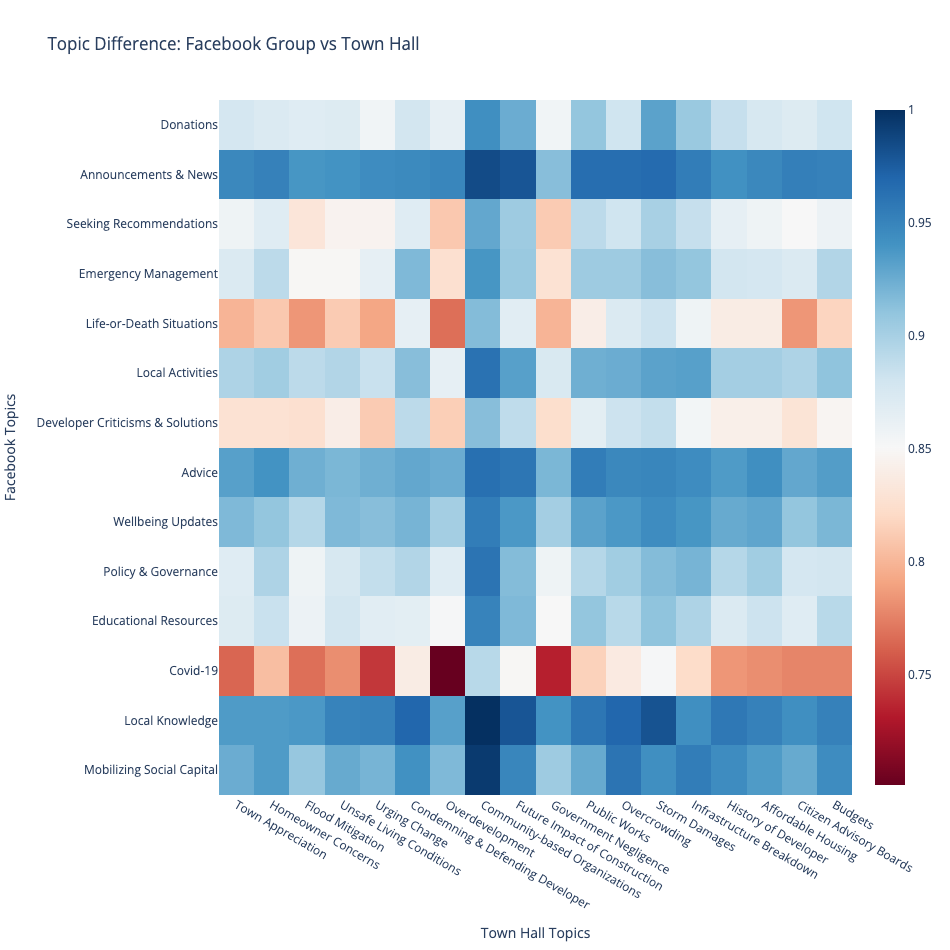}
  \caption{Heatmap of similarity between topics in the Town Hall Meetings and online neighborhood threads}
  \Description{Heatmap of Jaccard results}
  \label{fig:heatmap}
\end{figure*}

Figure \ref{fig:heatmap} shows a heat map illustrating the Jaccard similarity scores between topics in the Town Hall Meetings and online neighborhood forums. The blue to dark blue cells represent the strongest correlation between topics (0.90 <= J <= 1) and the orange to dark red cells indicate the weaker correlations (0.70 <= J <= 0.80) relative to the other topics. Overall, the online neighborhood forum and town hall meeting topics are moderate to highly correlated (0.70 <= J <= 0.99). 

\subsubsection{Strongest similarities} \label{strongest_similarity}
In Town Hall Meetings, Topic TH8 (community-based organizations) had the most similarity across the spread of topics in the online neighborhood forum (0.89 <= J <= 0.99). This finding supports our qualitative finding and prior research on the need for inter-organizational collaboration to solve complex problems post-disaster. Topic TH8 shared the highest similarity (J\verb|~|.99) with Topic ONG1 (mobilizing social capital) and Topic ONG2 (local knowledge). The high similarity between Topic TH8 and Topic ONG1 suggested the prevalence of community-based organizations in disaster recovery. The high correlation between Topic TH8 and Topic ONG2 indicated that organizations become part of the communication surrounding local issues (e.g., homelessness, school bullying, renters' rights) in more goal-oriented settings such as town-hall meetings.

Topic ONG13 (announcements and news) in the online neighborhood forum is highly correlated across the spread of town hall topics (0.93 <= J <= 0.98) which indicated that similar announcements related to the town are being shared in both online and offline settings. In the online neighborhood forum, Topic ONG8 (developer criticisms and solutions) shared the least similarity with Topic TH6 (condemning and defending developer, J\verb|~|0.76) and Topic TH17 (citizen advisory boards, J\verb|~|0.78) in the Town Hall Meetings. Upon further investigation of the of comments, we noticed differences in the use of language to describe the developer, a key figure relevant to those topics. In the online neighborhood forum, members used informal language (e.g., slumlord, leach, scum) to describe the developer. In Town Hall Meetings, public members used more appropriate language and referred to the developer by their name or the company's name.

In the online neighborhood forum, Topic ONG10 (discussions about life and death) had lower similarity scores across the spread of town hall topics, with Topic TH7 (overdevelopment) being the lowest (J\verb|~|0.76). Members reflected on the who, when, and why of local incidents that resulted in death (e.g., house fire, flash flooding, pedestrian-car accident). This finding suggests that individuals communicated in the online forum to make sense of the traumatic events. In contrast, communication in the Town Hall Meetings raised awareness of life-or-death incidents (e.g., pedestrian-car accidents) and proposed solutions (e.g., more crossing guards), suggesting that the speakers were past the initial sensemaking phase. Topic ONG3 (COVID-19) in the Online Neighborhood Group  was the least correlated across the spread of topics in the Town Hall Meetings, with the weakest being Topic TH7 (overdevelopment, J\verb|~|0.70). This finding revealed that community members discussed concerns related to COVID-19 (e.g., finding test kits, mask mandates) more with each other in an online neighborhood forum than with elected officials in Town Hall Meetings. 

\subsubsection{Weakest similarities} \label{subsec:finding_weekest} The findings revealed several differences between settings. Topic ONG8 (developer criticisms and solutions) in the online neighborhood forum showed relatively low similarity with Topic TH6 (condemning and defending developer, J\verb|~|0.78) in the Town Hall Meetings. This could imply that discussions about developers and their accountability may be more contentious and varied in the online neighborhood forum as opposed to the Town Hall Meetings. Additionally, Topic ONG10 (life-or-death situations) in the online neighborhood forum exhibited lower similarity scores across the town hall topics, with the lowest being Topic TH7 (balancing economic growth with community values, J\verb|~|0.76). This might suggest that expressions of empathy and personal concerns are more frequent in the online neighborhood forum than in Town Hall Meetings. Lastly, Topic ONG3 (COVID-19 testing) in the online neighborhood forum exhibited the weakest similarity across the spread of town hall topics. This finding suggests that the discussions related to COVID-19 testing are not as closely linked to the broader range of topics addressed in Town Hall Meetings.

\label{sec:findings}

\section{DISCUSSION} 
In this section, we reflect on our findings to discuss how the affordances of the Online Neighborhood Group  affected the discourse when compared to Town Hall Meetings. We then present a number of design recommendations based on our findings that might lead to the bridging of online and offline modalities of civic publics. Finally, we provide recommendations for governing leaders for more effective civic discourse. 

\subsection{Affordances of Physical and Digital Publics}
While there is a high similarity between discussion topics, we observed significant differences in the design of both contexts that can influence how citizens engage in communicative civic participation. Town Hall Meetings are designed to be speaker-focused, with citizens addressing the council during the public discussion portion of the meeting. This format does not encourage back-and-forth discussion among citizens, but rather enables them to raise concerns or ask questions directed at a more formal governing board \cite{field2019town}. Unlike in online settings, citizens cannot interject while someone is addressing the council, which may limit the depth of understanding and engagement with the issues under discussion. Although these meetings serve a vital democratic function, we contend that they are not conducive to the interactive, informal political discourse possible in online neighborhood forums.

\subsubsection{Back stage organizing}
In contrast to the formality of Town Hall Meetings, online neighborhood forums provide a setting for residents to engage in more open and candid political conversations. Consider Topic ONG8 (developer criticism and accountability), in which a resident used the term "slumlord" to describe a regional real estate investor. Instances of informal language like this suggest that users feel comfortable sharing their concerns and criticisms, indicating a sense of trust and camaraderie. Interestingly, Topic ONG8 in the online corpus yielded low Jaccard similarity scores across the range of topics in Town Hall Meetings addressing the similar issues (TH4, TH6, TH7). This observation can be ascribed to the structure of Town Hall Meetings, where participants tend to adopt a more formal tone in the presence of community leaders, government officials, and other stakeholders.

The online forum allows community members to express grievances using stronger language without fear of immediate confrontation or repercussions. Town Hall Meetings, on the other hand, promote the use of more restrained language, which may limit the depth and openness of discussions. In this context, the online neighborhood platform serves as a back stage \cite{goffman1959presentation}, facilitating the informal discussions necessary for delving deeper into complex community issues. Meanwhile, Town Hall Meetings function as a front stage where residents present their concerns to elected officials. This study underscores the importance of communities having diverse venues for communication, enabling citizens to engage in civic discussions more effectively. 

Herdağdelen et al. \cite[P.360]{herdaugdelen2023geography} argue that Facebook groups in densely populated areas are ``Bridging Groups...which bring together people [in the community] not already bound by friendship ties.'' Relatedly, in \S\ref{strongest_similarity} we found that TH8 (community-based organizations) is highly similar to ONG1 (mobilizing social capital) and ONG2 (local knowledge). Additionally, we know that 25 Facebook group users also contribute to Town Hall Meetings. This means that discourse about local organizations and local knowledge are being shared between community members who might otherwise not be connected in the community. 

\subsubsection{Organizing for Acute Events}  Iyengar et al. \cite{iyengar2024resilience} argue that communities can become more resilient if they use technologies designed to support people when they face acute stressors (e.g., flooding or a pandemic surge with high rates of infection). The structure of communication can also be attributed to affordances like synchronicity. We find that the online forum provides a more immediate response to community concerns compared to traditional venues for civic participation, such as Town Hall Meetings. Contrary to Hampton’s \cite{hampton2016persistent} assertion that the asynchronous nature of neighborhood forums reduces the pressure to formulate immediate responses, we argue that the immediacy of feedback in online forums depends on the urgency of community issues. While it is true that users can contribute to discussions at their own convenience, this flexibility can be overridden when there is significant pressure from pressing community concerns.

During disruptive events, the need for citizens and community organizations to mobilize social capital \cite{doerfel2017story} creates a sense of urgency that permeates online forums \cite{stephens2022social}. In these scenarios, the asynchronous nature of online forums does not necessarily lead to more casual communication, but rather, fosters a dynamic and responsive environment where community members can swiftly address urgent concerns. In contrast, Town Hall Meetings typically involve a more extended decision-making and implementation process before any action can be taken. Consequently, this longer timeline for addressing community issues can result in reduced responsiveness.

Recall that Topic ONG3 in the online forum (COVID-19) had the lowest Jaccard similarity scores across the range of topics in the Town Hall Meetings (see \S\ref{subsec:finding_weekest}). In this example, citizens turned to the online forum for immediate feedback on needs like finding COVID-19 tests during a national shortage. We also found that social support in the early acute COVID phases was reflected in the Online Neighborhood Group  interactions (see \S\ref{subsec:results_social_sup}). We argue that the immediacy of communication in neighborhood forums not only enables quicker responses to pressing concerns but also enhances the overall effectiveness of civic participation. By allowing community members to address issues in real-time, online neighborhood forums help to mobilize resources, generate ideas, and foster collaboration more efficiently than traditional venues for civic participation. This capacity echoes earlier work that shows how local groups organized to cope with the acute crises (e.g., early COVID surges) as presented in \S\ref{related_work_subsec:platform_posibilities} and \S\ref{related_work_subsec:local_communities_COVID}. 

\subsubsection{Organizing for Chronic Events}
Just as supporting communities facing acute stressors is important to the resilience and health of a community, it is equally important to provide local communities with support when facing ``long-term (chronic), recurrent, and persistent stressors.'' \cite{iyengar2024resilience} In \S\ref{subsec:results_recovery}, we showed how some forms of recovery can take on long-term properties. For example, ONG14 focused on sharing donations that allowed families to feel that there is a return to a normal. These were not acute needs like water or food, but children's toys and other gifts that might alleviate family stresses. The use of Facebook groups as local gift (donation) economies is in line with the categorization of Facebook groups in urban spaces as suggested by Herdağdelen et al. \cite{herdaugdelen2023geography} in \S\ref{related_work_subsec:platform_posibilities}. However, discourse about chronic stressors was not not always positive in nature as it extended to the discussion of root causes of societal problems like the lack of affordable housing. This in essence led to the ``long-now of [digital] community response to disaster'' \cite[P.13]{soden_et_al_21} (described in \S\ref{related_work_subsec:local_communities_COVID}) as the local community members discussed organizing around long-term advocacy goals (see \S\ref{subsubsec:comm_org}).

\subsection{Design Recommendation: Bridging Offline and Online Community Settings}
In light of this paper's findings, we recommend that designers better integrate and unify information from different \textit{offline} settings into digital community systems. A unified community interface can help facilitate interaction between Town Hall Meetings, either by streaming these events LIVE or archiving the videos from other sources (e.g., YouTube, the borough's website). If not recorded, platforms can consider merging information from Town Hall Meetings, which, along with any videos, are considered information open to the public domain. This approach has the potential to encourage broader participation by accommodating individuals regardless of their location and can help to enrich community dialogue by allowing real-time, synchronous interaction across different mediums \cite{albrechtslund2008online, willems2019politics}. If live streaming Town Hall Meetings, we recommend interactive tools like live polling and real-time commenting, which can help ensure that users are able to contribute actively to ongoing community discussions.

We also recommend embedding contextual awareness features into community-based platforms that can help alert users to discussions of interest based on their previous activities or expressed preferences. For example, when topics from past Town Hall Meetings are brought up online, participants who engaged previously could receive targeted notifications. This not only keeps the community engaged but also ensures continuity in discussions across platforms, fostering a deeper connection to local issues \cite{lopez2015lend, erete2015engaging}. Such a system would allow users to experience ``diffracted connections with groups'' discussing similar topics \cite[P.11]{lee_et_al_chrystal_22}  by ``transport[ing]'' them through the narratives of others facing similar experiences \cite[P.8]{randazzotrauma}.

\subsection{Policy Recommendation: Creating and Endorsing Digital Town Halls}
The results from this study highlight the significant need to cultivate spaces for community discussion in the wake of a disaster. We urge policymakers, community leaders, and disaster response organizations to prioritize the establishment of communication infrastructures such as online neighborhood forums. Our findings indicate that endorsing such platforms can improve the diversity of concerns raised, especially related to individuals most affected by disasters. This collaborative approach to civic engagement can contribute to community resilience by fostering the development of adaptive strategies \cite{doerfel2022resilience} and interventions that alleviate the impacts of disruptive events \cite{harris2022hyperlocal, doerfel2022resilience}. We recommend that municipalities create online neighborhood forums to foster open dialogue among residents. However, we advise against a strong or prominent presence of elected officials in these online spaces. Instead, we propose enlisting local volunteers to serve as community moderators, acting as brokers of information between residents and government representatives. This approach can help maintain the informal and unrestricted nature of discussions while ensuring that citizens do not attempt to resolve issues in isolation. Forum moderators can help communicate essential information to citizens, which allows for a degree of organizational involvement in the online forum. Ultimately, this strategy can serve to reinforce democratic processes by offering a platform for citizens to express their concerns and opinions while also promoting transparency and accountability within local government.

\section{Limitations and Future Work}
Our analysis in this paper focused on discourse across offline and online contexts in civic interaction. Another limitation in this analysis is the lack of temporal analysis of the discourse in both modalities over time. Such an analysis can provide support to our findings that suggest Online Neighborhood Group s provide support in more time-sensitive conditions. It will also show if and how discussions in one modality affect the other over time. A related avenue is a deeper analysis of user networks that compares users who cross boundaries of online and offline modalities with those who do not and how they relate to other members of the community. Future research can also examine a wider range of civic participation scenarios to explore how the dynamics of online and offline engagement might vary across different settings or situations. 

Another limitation of this study is that it focuses on one community using one social media platform and one disaster. While we think that many of the findings can be adapted into the design of online communities for better online disaster organizing, future work should include a larger number of communities, and a larger number of online platforms that can provide a wider variety of affordances to community members. Since the needs of online community organizers are different in urban and rural spaces \cite{birbak_FB_groups_snow_2012,herdaugdelen2023geography}, and since different platforms present users with different affordances (as reviewed in \S\ref{related_work_subsec:platform_posibilities}), future work should compare a variety of configurations that cover a matrix of challenges and opportunities for the support of communities through crises.

\section{CONCLUSION}
This study sheds light on the complementarity of online and offline civic participation following an extreme weather event. Our analysis reveals that both online neighborhood forums and Town Hall Meetings contribute to meaningful discussions, challenging previous assumptions regarding the quality of discourse in digital spaces. Our findings emphasize the value of online neighborhood forums as a platform for fostering open dialogue, support networks, and collective action, and highlight the importance of providing diverse venues for civic communication. This study primarily focuses on the content of civic discussions and the environments in which this discourse is situated rather than the social networks underpinning civic interactions. Future research could use social network methods to examine the structure, composition, and dynamics of social ties among community members and assess how these relationships influence individual and collective engagement in democratic processes.

\bibliographystyle{ACM-Reference-Format}
\bibliography{software}

\end{document}